\documentclass[12pt]{article}
\usepackage{latexsym,amsmath,amssymb,amsbsy,graphicx}
\usepackage[cp1251]{inputenc}
\usepackage[T2A]{fontenc}
\usepackage[russian,english]{babel}
\usepackage[space]{cite} 

\begin{document}
\bigskip

{\bf
FLARE ACTIVITY OF THE SUN AND VARIATIONS IN ITS UV
EMISSION DURING CYCLE 24
}

\bigskip

\centerline {E.A. Bruevich, G.V. Yakunina}

\centerline {\it Lomonosov Moscow State
 University, Sternberg Astronomical Institute,}
\centerline {\it Universitetsky pr., 13, Moscow 119992, Russia}\

\centerline {\it e-mail:  {red-field@yandex.ru, yakunina@sai.msu.ru}
 }\

\bigskip

{\bf Abstract.} The flare activity and the ultraviolet emission of the sun during its 24-th cycle are analysed. As compared to
cycles 21-23, where the most powerful flares were observed during the decay phase, in cycle 24 the greatest number of 
powerful flares (>X2.7) occurred in the rising phase and at the maximum with the exception of the two largest flares of cycle 24 X9.3 and X8.2 in September 2017.
We showed that regression fits of solar UV
indices to the overall radiation level from the sun are substantially different for cycle 24 compared to
cycles 21-23. It is found that for the flare of August 9, 2011 (SDO and GOES-15 observations), the flare
propagates in a direction from the upper corona to the transition region and to the chromosphere. A study of the
N-S asymmetry in the distribution of the flares in cycle 24 reveals a strong predominance of flares in the
N-hemisphere in 2011 and in the S-hemisphere in 2014. It is also found that during cycles 23 and 24, the
delays in the onset of proton events relative to the onset of the flares that cause them have a distribution
with a distinct maximum corresponding to a delay of 2 hours for protons with energies >10 MeV, as well
as for those with energies >100 MeV.

\bigskip
{\it Key words.} Sun: cycle 24: flares: flare index: variations in UV emission: proton events.
\bigskip

\vskip12pt
\section{Introduction}
\vskip12pt

 Analyses of the flare activity and of variations in the UV emission of the sun are important for research on
solar and terrestrial physics. The accelerated particles generated by powerful flares have a fundamental influence on
space weather and have many practical applications relating to the study of physical processes in the heliosphere
and in the earth's atmosphere.

\vskip12pt
\section{Features of activity cycle 24}
\vskip12pt

The current solar activity cycle (No. 24) is the weakest solar cycle over
the last 100 years or more. At present the transition to the following activity minimum between cycles 24 and 25, which is expected in roughly 2018-2019, is taking place. 

Solar cycle 24 continues the trend of recent years of a
reduction in the number of sunspots since cycle 21, which reached its maximum in roughly 1980 (see Fig. 1). Many scientists at NASA who are involved in the prediction of solar activity assume that cycle 25 will be roughly the same or weaker than 24. The variations in the solar activity have been modelled numerically [1] using observational data
on the number of sunspots in the period 1750-2050. 

The dynamo theory was used to solve the evolution equation
for a number of relative sunspot number (SSN) with the mechanism for magnetic field formation in sunspots taken into account. The
model results [1] correlate well with observations and predict a prolonged period of low activity up to 2050 similar to the Dalton minimum.

Fig. 1 shows the variations in the number of sunspots over 300 years. The current cycle 24 is clearly one of the lowest and is comparable to the activity during the epoch of the Dalton minimum. It is expected that cycles
25 and 26 will be similar to cycle 24.

\begin{figure}[tbh!]
\centerline{
\includegraphics[width=120mm]{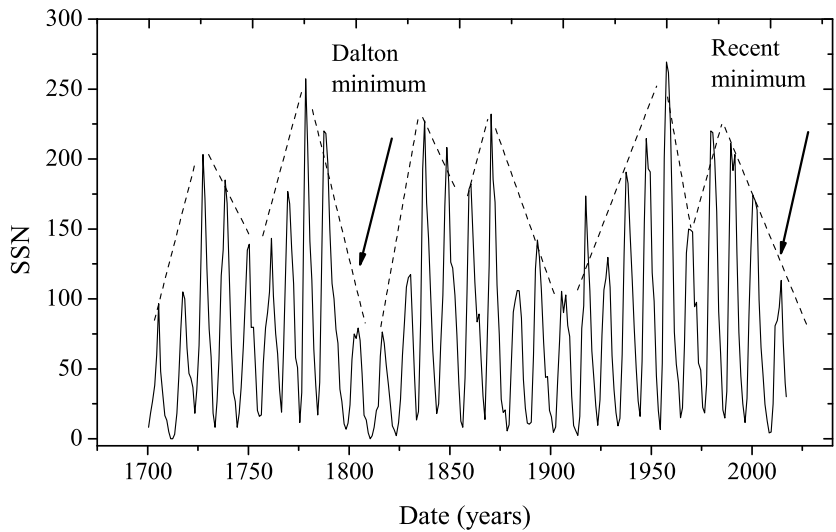}}
 \caption{Yearly average number of sunspots (SSN) from 1700 to the present.}
{\label{Fi:Fig1}}
\end{figure}

\vskip12pt
\subsection{Solar activity in the ultraviolet range and flare activity during cycle 24}
\vskip12pt 

Studies in the ultraviolet (UV) are an important part of research on the development of solar flares. The UV part of the electromagnetic spectrum covers the range from 5 to 400 nm. UV photons are absorbed in the upper layers of the  atmosphere; short-wavelength
fluxes of solar radiation cause ionization and dissociation of atmospheric species and lead to formation of the ionosphere. The solar UV emission is formed in the upper chromosphere and the transition region, while X-rays are
formed in the corona; the fluxes in these ranges make up a comparatively small fraction of the overall radiative flux.

As opposed to most stars of the sun's type, the sun is characterized by comparatively low spottedness, moderate flare
activity, and a low level of corona emission flux [2,3]. 

\begin{figure}[tbh!]
\centerline{
\includegraphics[width=100mm]{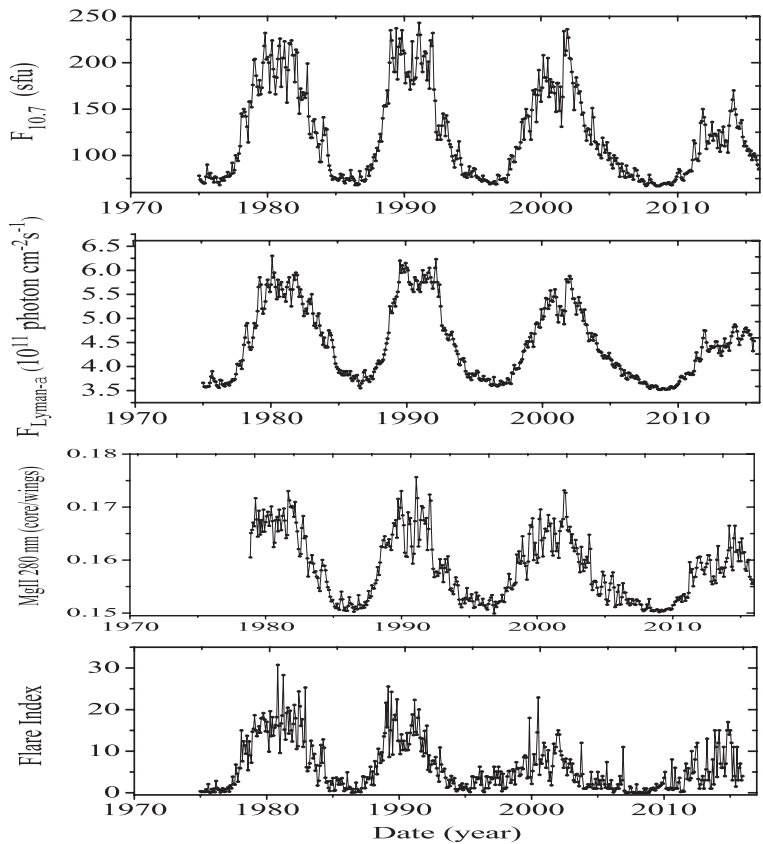}}
 \caption{Cyclical variations in the fluxes $F_{10.7}$, $F_{Ly -\alpha}$, and MgII 280 nm
(core/wings) and the flare index (FI) observed during cycles 21-24.}
{\label{Fi:Fig2}}
\end{figure}

Against this background, even very small variations in the fluxes associated with the appearance and disappearance of groups of spots in the active regions or variations in
activity during the solar cycle and large flares can cause significant changes in the UV and X-ray activity indices.

The UV fluxes analysed in this paper vary by tens of percent on time scales ranging from a few minutes to several hours (during solar flares); by a few percent over times ranging from a few days to several months (owing to the movement of active regions over the disk because of the sun's rotation); and by as much as a factor of two over times
ranging from a few years to decades (variations in the solar cycle). This variability produces variations in the earth's ionosphere and upper atmosphere that are similar in amplitude and time. Since the solar UV emission is mainly absorbed in the upper layers of the atmosphere, it is measured using instruments mounted on rockets and satellites.

Because of the related difficulties, for many years different predictions of the UV fluxes were used that employed correlations with the numbers of sunspots and 10.7 cm radio wave emission ($F_{10.7}$). The $F_{10.7}$ flux is now often used
as an objective index for the current level of solar activity; this flux is measured in sfu (solar flux units), with 1 sfu corresponding to a flux of $10^{-22}$ $W/m^2/Hz$. Thus, a comprehensive analysis of observational data on solar UV emission
is needed, since the emission in this region of the sun's spectrum is the most important input parameter for modelling the state of the earth's ionosphere and determines the space weather in the earth's vicinity.

It has been shown [4] that the relationship between the UV solar activity index and the general activity of the sun based on the value of $F_{10.7}$ differed for cycles 21-23, with changes in the magnitudes of the regression coefficients. The correlation coefficient for a linear regression varies within a single cycle, with two maxima during
the rising and decreasing phases and two minima during the cycle minimum and maximum [4].

Fig. 2 shows time series of the ultraviolet activity indices and the flare index compared with a series of $F_{10.7}$ as indicators of the overall activity of the sun's atmosphere.
Fig. 2 shows that the current cycle 24 has
amplitudes of the UV indices and the flare index that are roughly half those for cycles 21-23. This corresponds to the lower level of solar activity as a whole at the present time.

The goals of this paper are:

(1) to analyse the variations in the UV Lyman alpha 121.7 nm and MgII 280 nm lines, as well as of the flare activity
of the sun in cycle 24;

(2) to compare the fluxes at wavelengths ranging from soft X-ray (data from GOES satellites for 0.1-0.8 nm) to
ultraviolet (data on fluxes in five UV lines from the Solar Dynamics Observatory - SDO) in order to track the
propagation of the largest flare in cycle 24 in the solar atmosphere;

(3) to estimate the delay in the onset of proton events relative to the onset of flares in the 0.1-0.8 nm range associated with these events based on data from the GOES 13-15 satellites.

\vskip12pt
\section{UV fluxes and flare activity in cycle 24}
\vskip12pt

\vskip12pt
\subsection{UV radiation flux in the MgII 280 nm line}
\vskip12pt

The index characterizing the ratio of the fluxes in the center
and wings of the atmospheric MgII (280 nm) line has been observed continuously using instruments on satellites since 1978. The core of the doublet MgII 279.56 and 280.27 nm lines is formed in the chromosphere, but the wings are
mainly of photospheric origin. This index is a good indicator of the state of the solar atmosphere. 

A close coupling is observed [5,6] between it and the flux in different lines and ranges of UV emission from the sun, which are important
for modelling the state of the earth's atmosphere. The index MgII (core/wings) is measured in relative units. The
observations from Ref. 7 are used.

Fig. 3 shows the MgII 280 nm (c/w) UV radiation flux as a function of the radio emission flux at a wavelength of 10.7 cm, an indicator of the level of solar activity. The solid points show the dependence for cycles
21-23 and the hollow circles, the dependence for cycle 24. Separate quadratic regression fits are shown for cycles
21-23 and for cycle 24. These fits differ significantly: in cycle 24 the maximum fluxes were roughly half the
magnitude of the maxima in cycles 21-23, the spread in the values is larger in cycle 24, and the regression coefficients
differ substantially.

\begin{figure}[tbh!]
\centerline{
\includegraphics[width=100mm]{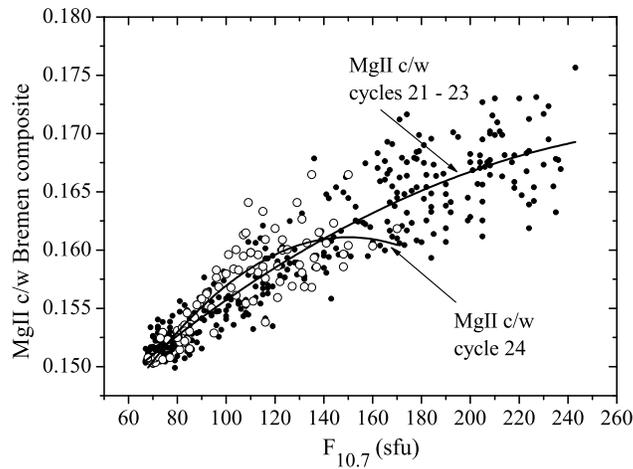}}
 \caption{MgII c/w  as a function of
the radio emission flux at 10.7 cm}
{\label{Fi:Fig}}
\end{figure}

\vskip12pt
\subsection{The UV flux in the Lyman $ \alpha $ 121.6 nm line}
\vskip12pt

The flux in the Lyman-alpha hydrogen line ($F_{Ly -\alpha}$) from the entire
solar disk is an important indicator that characterizes the state of the chromosphere and the lower part of the transition region [8]. The index $F_{Ly -\alpha}$ is measured in units of $10^{11} photons/cm^2/s$. The observations from Ref. 9 are used.

Fig. 4 shows the dependence of the ultraviolet emission $F_{Ly -\alpha}$
at 121.6 nm on the 10.7 cm radio emission.
The solid points show this dependence for cycles 21-23 and the hollow triangles, the dependence of cycle 24.

Quadratic regression fits are shown here: the regression curves and fits for these two dependences differ more distinctly
than for the case of the MgII 280 nm c/w index.

\begin{figure}[tbh!]
\centerline{
\includegraphics[width=100mm]{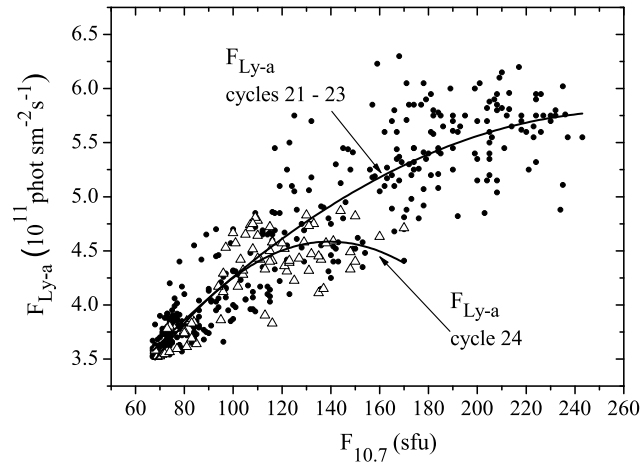}}
 \caption{The $F_{Ly -\alpha}$ 121.6 nm UV radiation flux as a function of
the radio emission flux at 10.7 cm}
{\label{Fi:Fig}}
\end{figure}

\vskip12pt
\subsection{The flare index in cycle 24}
\vskip12pt

The distinctive feature of the time variations in the flare index at the end
of the 20th and beginning of the 21st centuries is that by cycle 23 the maximum of FI had fallen by a factor of two compared to cycles 21 and 22. Archived data on FI from 1975 to 2008 are available at 
$$http://www.ngdc.noaa.gov/
stp/space-weather/solar-data/solar-indices$$
For cycle 24 we calculated FI from Ref. 10 using data on flare activity that are available at 
$$http://www.wdcb.ru/
stp/data/Solar_Flare_Events/Fl_XXIV.pdf$$ 
(see Fig. 2). It was found that the amplitude of FI during cycle 24 decreased
by a further factor of 2 compared to cycle 23.

\begin{figure}[tbh!]
\centerline{
\includegraphics[width=100mm]{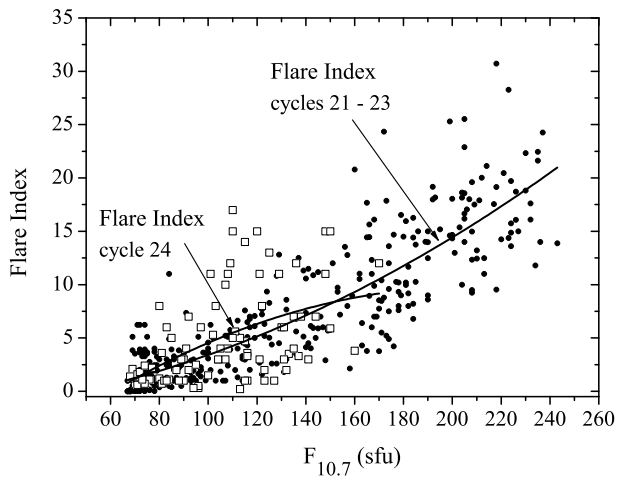}}
 \caption{The flare index FI as a function of the radio emission
flux at 10.7 cm}
{\label{Fi:Fig}}
\end{figure}

Fig. 5 shows the dependence of FI on the radio emission flux at 10.7 cm. The solid points are for cycles 21-23 and the solid squares are for cycle 24. Quadratic fit curves are shown: the differences in the regression curves for the two dependences are not as marked as in the case of the UV indices, but the spread in the values is greater.

\vskip12pt
\section{ Large flares in cycle 24}
\vskip12pt

Fig. 6 shows the distribution of the MI-X7 (740) flares in cycle 24. The points indicate the months for which the number of flares per month was N<6; the asterisks, the months for which 5 < N < 10; and the large hollow circles, the months for which N > 20. 

The data for the flares are plotted on a curve showing the monthly-mean numbers of solar spots. It is clear that the largest number of flares took place in the falling branch after the first maximum,
and during the second maximum and the falling branch of cycle 24. The largest number of large flares occurred
in 2014. Before 2011 and since mid-2015, almost no large flares have been observed. As compared to
cycles 21-23, where the most powerful flares were observed during the decay phase, in cycle 24 the greatest number of 
powerful flares (>X2.7) occurred in the rising phase and at the maximum with the exception of the two largest flares of cycle 24, which occurred almost at the minimum between 24 and 25 cycle in September 2017.

\vskip12pt
\section{North-south (N-S) asymmetry in the localization of large flares in cycle 24}
\vskip12pt

\begin{figure}[tbh!]
\centerline{
\includegraphics[width=100mm]{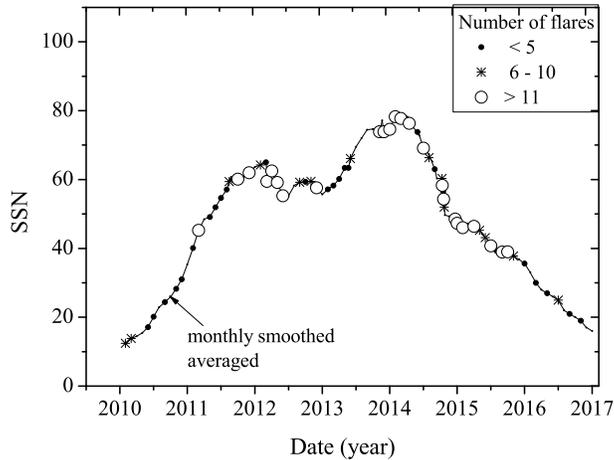}}
 \caption{Monthly averaged distribution of large flares
during cycle 24}
{\label{Fi:Fig}}
\end{figure}

The basic features of the N-S asymmetry have been studied and described in recent years. These show up in all the indices of solar activity, but the reason for this asymmetry is not yet understood. Probably the N-S asymmetry in the indices is a fundamental property of solar activity that is indicative of significant differences in processes taking place in the northern and southern hemispheres. In a study of the N-S asymmetry [11] it was found that many features of the cyclical activity of the sun show up more strongly in the asymmetry of the activity indices
than in the values of these indices as such.

An analysis of solar activity in cycles 21-23 [12] showed that the observed large flares in the soft X-ray range (GOES, 0.1-0.8 nm) in different phases of solar cycles 21-23 had an N-S asymmetry. It was also found that over the entire cycle 21, a small excess of flare activity was observed in the northern hemisphere, while during cycles 22 and 23, a southern excess predominated. It is important to note that an N-S asymmetry of solar flares shows
up both in the number of flares and in the intensities of the flares; this should be taken into account in models of the solar dynamo [12].

We have studied the N-S distribution of flares during cycle 24. 

Table 1 lists data from the analysis of the N-S asymmetry for large flares of X-ray class > M5 (119) in different phases of cycle 24. It is clear that a strong asymmetry in the flare activity was observed during 2011 (15 flares in the northern hemisphere and 4 in the southern) and 2014 (11 in the northern hemisphere versus 28 in the southern). During the rest of cycle 24, the number of flares in the two hemispheres were roughly equal.

Fig. 7 illustrates the N-S distribution of large flares (monthly average data for X-ray class > M1) in different phases of cycle 24.
On the plot of the sliding monthly average numbers of sunspots, different symbols are used to denote the months with distinct N-S asymmetries. In this figure, the solid circles indicate the years when the number of flares in the northern hemisphere is substantially higher than in the southern hemisphere, the asterisks indicate predominant flares in the southern hemisphere, and the hollow circles indicate roughly equal number of flares in the two hemispheres.

\begin{figure}[tbh!]
\centerline{
\includegraphics[width=100mm]{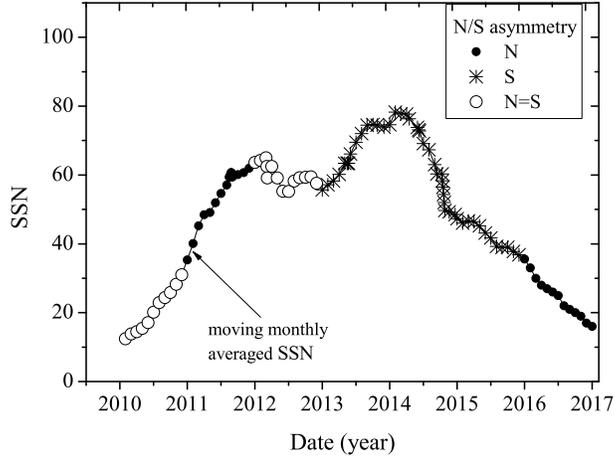}}
 \caption{The N-S asymmetry in the number of large
(>M1) flares during cycle 24}
{\label{Fi:Fig}}
\end{figure}

It can be seen from Fig. 7 that in the rising phase of cycle 24 the numbers of large flares are initially roughly equal in the northern and southern hemispheres. Then in 2011 and 2012 the number of flares is seen to be larger in the northern hemisphere; this corresponds to an asymmetry in the number of spots. 

During the time of the first maximum, a roughly equal number of flares is observed in the two hemispheres. During the second maximum, in the decreasing phase of the cycle from 2013 to 2016, we see that flares predominate in the southern hemisphere;
this corresponds to the asymmetry of sunspots during cycles 21-23. In 2016, we again see more flares in the northern hemisphere.

An analysis of the rotation velocities of sunspots and the evolution of the active longitudes [13] shows that the northern and southern hemispheres rotate with different velocities and that the rotation velocities in both hemispheres have increased since 1990. It is assumed that the observed north-south asymmetries of groups of sunspots and flare activity are consequences of the temporal evolution of the large-scale magnetic field and the solar dynamo.

It is clear that the asymmetry in the flare activity correlates well with the asymmetry in the appearance of groups
of sunspots according to data in the archives of the Solar Influences Data Analysis Center (SIDC) of the Royal Observatory of Belgium, which includes the World Data Center for the study of sunspots. 

According to these data, for cycles 21-23 and the rising phase of cycle 24, there is an N-S asymmetry in the number of spots that differs slightly from the N-S asymmetry in the flare activity; for all of cycles 21-24 the number of spots in the northern hemisphere
predominates in the rising phases of the cycles, while the number of spots in the southern hemisphere predominates in the decreasing phases; see 
$$http://www.sidc.be/silso/datafiles$$ 
The observations of N-S asymmetry in the various
indices are of great significance, since they provide valuable information for modelling the solar dynamo. The appearance of a substantial southern asymmetry at the start of the 21st century indicates the onset of a period of small
activity cycles [14], which we are observing at the present time.

\vskip12pt
\section {Analysis of satellite observations of the X6.9 flare of August 9, 2011}
\vskip12pt

\begin{center}
\begin{table}
\caption{The Asymmetry in Flare Activity During Cycle 24}
\vskip12pt
\begin{tabular}{||c|c|c|c|c||}

\hline \hline

  Year &Number flares&Number flares &Number flares&Number flares\\

     &     > M5   &   > X1     &   in N hemisph. &  in S hemisph. \\  \hline               
  
2010 &  3   &   0 & 2 &  1 \\ \hline

2011 &   19   &  8   &  15 & 4  \\ \hline

 2012 &    21  &  6   & 12  &  9 \\ \hline
 
2013   &   21   &  12   &  10 & 11 \\ \hline

 2014 &  39    &  16   & 11  &  28 \\ \hline

2015 &  12    &   2  &   5  &  7  \\ \hline

2016 &  4    &  -   &  4   & - \\ \hline

\hline \hline 
\end{tabular}
\end{table}
\end{center}

\begin{figure}[tbh!]
\centerline{
\includegraphics[width=100mm]{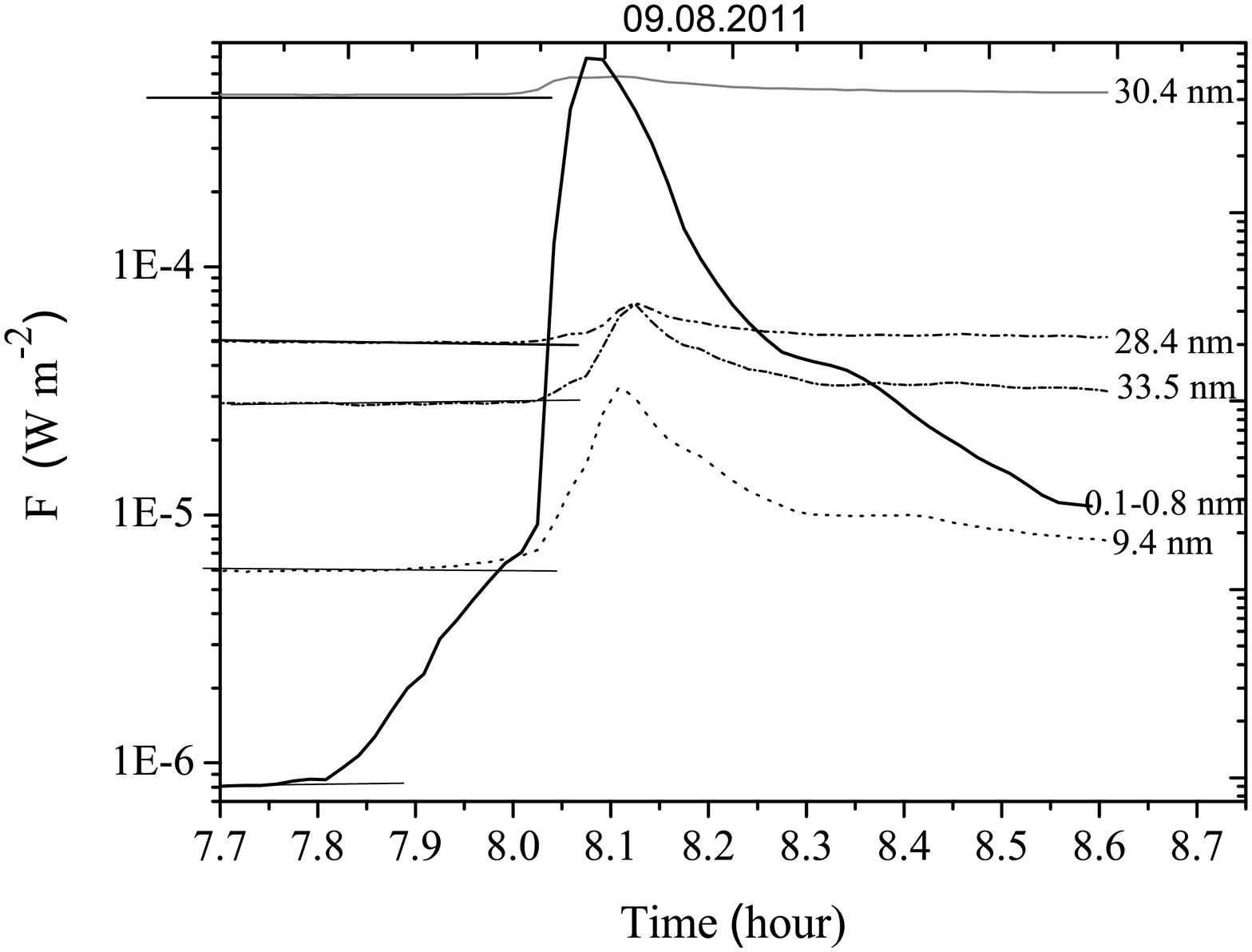}}
\caption{The evolution of the flare of August 9, 2011, in
different UV lines and in the X-ray range (the 30.4, 33.5,
28.4, and 9.4 nm lines and the 0.1-0.8 nm range)}
{\label{Fi:Fig}}
\end{figure}

UV fluxes have been measured with high accuracy by satellites since 1999. This has made it possible for us to analyse the onset times in different lines of the largest flare during cycle 24, X 6.9, which took place on August
9, 2011 using observational data from SDO/EVE and GOES-15. 

This is important for refining the models of flares.
We note the times associated with flare intensification of the fluxes in coronal lines and in lines formed in the chromosphere and transition region. The regions of the coronal flare plasma in which UV and X-ray emission is formed have different heights and temperatures. The gradual increase in the fluxes in these lines during the development of a flare can be used to determine the temperatures within the regions where the flare begins, as well as the in the regions where it propagates.

Fig. 8 shows the evolution of the flare of August 9, 2011, in the coronal lines in the ranges of 9.4, 33.5, and 0.1-0.8 nm, as well as in lines (30.4 and 28.4 nm) of the chromosphere and transition region. The wavelengths are indicated on the right hand side of the figure. 

The horizontal smooth lines indicate the level of the background radiation flux; this makes it possible to estimate more accurately the time when the fluxes begin to increase. Note
that for shorter wavelengths of a line or spectral range, the ratio of the maximum flare amplitude to the background level is higher. It can be seen that the rise in the fluxes begins gradually (pre-flare enhancement). 

In the 0.1-0.8 nm range and the 9.4 nm line, the pre-flare begins at 7h 48m and 7h 54m, while the pre-flare (which is less distinct) in the other lines begins at 8h 00m. 

Thus, the pre-flare enhancement propagates from the higher levels of the corona into
the lower corona and chromosphere. We see the onset of the flare as a sharp, almost vertical increase in the flux
for 0.1-0.8 nm and a steep rise for 9.4 nm at 8h 02m, while the onset of the flare for the other lines occurs a few minutes
later at 8h 04m-8h 05m.

\vskip12pt
\section{Characteristics of the proton events in cycle 24}
\vskip12pt

Solar proton events (SPE) and the flares accompanying them during cycle 24 were analysed on the basis of
data from GOES 13-15. The most powerful SPE with proton energies E>10 MeV and E>100 MeV were identified
and studied.

SPE are observed in interplanetary space as a result of flares on the sun, as a result of ejection of coronal
mass, and sometimes after the disappearance of filaments. 

The duration of a proton event depends on the energy
of the protons. For energies ~10 MeV the duration is a few hours; for higher energies they can last up to several days. 

At present there are no unique answers to the question of where, when, and how the protons are accelerated and how they propagate to the observer. The power of proton events is measured in 

pfu ($particle flux units, protons/
cm^2/s/sr$).

\begin{center}
\begin{table}
\caption{Flares in Cycle 24 Accompanied by the Largest-scale Proton Events}
\vskip12pt
\begin{tabular}{||c|c|c|c|c||}

\hline \hline

  Date &X-ray/optic scale&Position &Proton flux
& Proton flux  \\

     &       &   on the disc     &  (>10 MeV), pfu&  (>100 MeV), pfu\\  \hline               
  
04.08.11 & X2.3/2B  & N19W76 & 99 &  2\\ \hline

23.01.12&  M8.3/2B   &  N28W21   & 4500 & 3 \\ \hline

 27.01.12 & X1.7/2F   &  N27W71  & 800  &  11 \\ \hline
 
07.03.12  &  X5.4/2B   & N17E27   & 200 & 70 \\ \hline

07.03.12 &  X1.3/SF   &  N23E12   & 35 &  3 \\ \hline

13.03.12 &  M7.9/1B    &   N19W59 &   500  &  2  \\ \hline

17.05.12 &  M5.1/1F   &  N11W76  &  250  & 20 \\ \hline

12.07.12 &  X1.4/2B   &  N13W15  & 100 & 0.2\\ \hline

17.07.12 & M1.7/1F    &   S28W90  &  130  &  0.1  \\ \hline

19.07.12 &  M7.7/SF   &  S16W90   &  90  & 0.7 \\ \hline

11.04.13 &  M6.5/3B    &   N09E12  &   100  &  2  \\ \hline

22.05.13 & M5.0/3N   & N15W70  &  2000   & 4 \\ \hline

07.01.14 &  X1.2/2N    &   S15W11  &   1000 &  4 \\ \hline

10.09.14 &  X1.6/2B   &  N12E02  &  130   & 1\\ \hline

21.06.15 &  M3.8/2B    &   S18W57  &   1000  &  0.1  \\ \hline

06.09.17 &  X9.3/3B    &  S15W45   &  50   & 0.7 \\ \hline

10.09.17 &  X8.2/3B    &  S12W85   &  650   & 70 \\ \hline

\hline \hline 
\end{tabular}
\end{table}
\end{center}
\bigskip

\begin{figure}[tbh!]
\centerline{
\includegraphics[width=100mm]{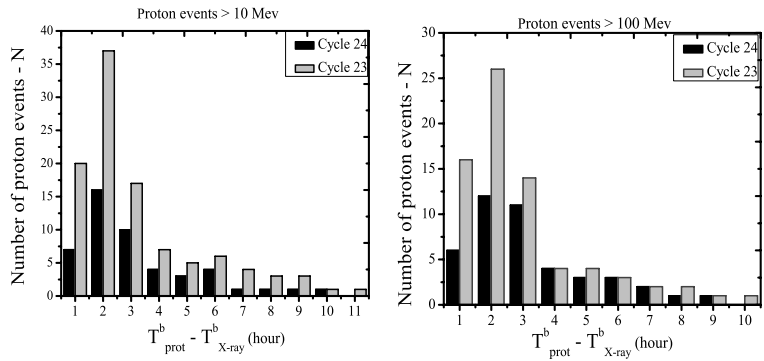}}
\caption{Histogram of the time delays of the start of proton events relative to the onset
of the flares that produce the events in the soft X-ray range}
{\label{Fi:Fig}}
\end{figure}
\bigskip

Proton events produced by flares and not by other causes (ejection of coronal mass, eruptive prominences) were selected on the basis of the observational data. Proton events in cycles 24 (black) and 23 (grey) were examined separately. The distributions of the delays between the flare onset in the X-ray range and the start of a proton event are similar in cycles 24 and 23. 

The maxima of the distributions occur for a delay of 2 hours, both for >10 MeV
protons and >100 MeV protons. Note that the most powerful flares are accompanied by proton events with the shortest delays. The largest delays are observed for proton flares lying in the eastern hemisphere because of a longer trajectory for propagation of the protons from the flare region.

Cycle 24 was distinguished by the fact that no proton fluxes with energies >100 MeV and characterized by a maximum amplitude of more than 100 pfu were observed.
During activity cycle 24, the number of flares accompanied by proton fluxes was lower by roughly a factor of 2. 

The characteristics of the energetic solar protons in the first 64 months of cycle 24 (the rising phase and maximum) have been compared with those of the previous cycles (21-23) [15]. It was found that, despite the low solar activity during the rising phase and maximum of cycle 24, the number of SPE with proton energies E>10 MeV and E>100 Mev in the current cycle differs little from the number of these events in cycles 21-23. During the period from 2010-2016, we identified 62 SPE with proton energies E>10 MeV and 24 with E>100 MeV (the most powerful
of these are listed in Table 2). 

Some of the proton events (including the event of March 7, 2012) have a complicated
time profile with several maxima and have a high power (see Table 2).
It can be seen that a large fraction of these events originated in the northern hemisphere during 2011-2013, while two out of three of the large flares with proton fluxes during 2014-2015 occurred in the southern
hemisphere. 

It is clear from Table 2 that the largest proton events were associated with flares of class M5 or higher. For the large flares in classes M1-X7 accompanied by fluxes of protons with energies >10 MeV and >100 MeV, we have analysed the time delay between the onset of a flare in the X-ray range and the onset of the proton event caused by this flare (see Fig. 9). An estimate of the delay in the arrival of the protons in the earth's atmosphere is
important for predicting space weather and the state of the upper atmosphere.

\vskip12pt
\section{Conclusion}
\vskip12pt

It follows from this paper that the flare activity during cycle 24 was considerably lower than in cycles 21-23. As compared to cycles 21-23, where the most powerful flares were observed during the decay phase, in cycle 24 the greatest number of 
powerful flares (>X2.7) occurred in the rising phase and at the maximum with the exception of the two largest flares of cycle 24, which occurred almost at the minimum between 24 and 25 cycles in September 2017.

An analysis of the ultraviolet fluxes in the MgII 280 nm and La 121.6 nm lines showed that the regression fits as a function of the overall emission level of the sun differ substantially in cycle 24 and in cycles 21-23.

The evolution of the flare of August 9, 2011, as observed in coronal lines (9.4, 33.5, and 0.1-0.8 nm) and in lines of the chromosphere and transition region (30.4 and 28.4 nm) proceeds in a direction from the upper corona toward the lower corona and chromosphere.

A study of the N-S asymmetry in the flare distribution during cycle 24 showed that a strong N-asymmetry of the flares was observed in 2011 and an S-asymmetry in 2014. In the other years of cycle 24, the numbers of flares in the two hemispheres were roughly the same.

During cycle 24, proton events with E>100 MeV are less intense; this confirms the conclusions of Ref. 15. The largest-scale proton events were associated with X-ray flares of class >M5.

Histograms of the time delays of the onset of proton events relative to the onset in the soft X-ray range of the flares causing these events during cycles 24 and 23 show that the maximum of the distribution occurs for a delay of 2 hours for protons with energies >10 MeV, as well as for those with energies >100 MeV.

\end{document}